\documentclass{elsart41}
\usepackage{graphics}
\usepackage{graphicx}
\usepackage{amssymb}
\begin{document}

\begin{frontmatter}

% Title, authors and addresses

% use the thanksref command within \title, \author or \address for footnotes;
% use the corauthref command within \author for corresponding author footnotes;
% use the ead command for the email address,
% and the form \ead[url] for the home page:
% \title{Title\thanksref{label1}}
% \thanks[label1]{}
% \author{Name\corauthref{cor1}\thanksref{label2}}
% \ead{email address}
% \ead[url]{home page}
% \thanks[label2]{}
% \corauth[cor1]{}
% \address{Address\thanksref{label3}}
% \thanks[label3]{}

\title{Thermal transport properties of disordered spin-1/2 systems}
%---- Don't remove this comment line! ----
%
% use optional labels to link authors explicitly to addresses:
% \author[label1,label2]{}
% \address[label1]{}
% \address[label2]{}

\author[AA]{Fabian Heidrich-Meisner}
\ead{f.heidrich-meisner@tu-bs.de}

\address[AA]{Technische Universit\"at Braunschweig, Institut f\"ur Theoretische Physik,  Mendelssohnstrasse 3, D-38106 Braunschweig, Germany}

\corauth[Fabian Heidrich-Meisner]{Corresponding author. Tel: +49-531-3915184;
Fax: +49-531-3915833}

\begin{abstract}

This work studies  heat  transport of bond-disordered spin-1/2 chains.
As an example, the $XX$ case is analyzed, which 
corresponds to a model of noninteracting spinless fermions. Within the fermion representation,
the single-particle eigenenergies are determined numerically, which allow one to compute 
transport coefficients. Since the ballistic transport properties of a clean chain are destroyed by
disorder, the focus  is  on the frequency dependence of
the thermal conductivity and  on  a qualitative comparison
with the spin conductivity, both  at finite temperatures.

\end{abstract}

\end{frontmatter}

% main text

Motivated by recent experiments on transport properties 
of transition metal oxides \cite{exp}, intense theoretical work has recently been devoted to the 
study of heat conduction in one-dimensional spin-1/2 systems
such as the Heisenberg chain, frustrated chains, and spin ladders \cite{zotos_rev}.
%\cite{zotos97,alvarez02,kluemper02,hm02,saito03,orignac03,zotos04,rozhkov05}).
Among these systems, the anisotropic spin-1/2 chain exhibits ballistic
thermal transport properties at all temperatures $T$ due to its integrability \cite{zotos97}. 
The thermal conductivity diverges in the homogeneous case,
signaled by a finite thermal Drude weight \cite{xxz}. Here, the usual decomposition of  
the real part of the thermal conductivity $\kappa$ into the Drude
weight $D_{\mathrm{th}}$ and the regular part is utilized:
$\mbox{Re}\,\kappa(\omega)=D_{\mathrm{th}}\delta(\omega)+\kappa_{\mathrm{reg}}(\omega)$,
$\omega$ being the frequency.
Ballistic transport 
is expected to be destroyed by randomness in the exchange couplings,
resulting in a vanishing  Drude weight. 
Therefore, one is interested in the  regular part $\kappa_{\mathrm{reg}}(\omega)$, from which 
the dc-conductivity can be extracted by extrapolating 
to $\omega=0$.\\\indent
In this contribution, the effect of
bond disorder on the thermal conductivity of $XX$ chains is studied numerically.
 This limiting case of the anisotropic spin-1/2 chain corresponds to 
 free spinless
fermions by means of
a Jordan-Wigner transformation \cite{mahan}. While  particle transport in  disordered fermion systems 
is a long-studied problem,  closely related to the subject of localization (see, e.g., Ref.~\cite{kramer93}),
 the thermal conductivity $\kappa$ in these models has  attracted less attention. 
 Using bosonization, predictions were made for the concentration dependence of the thermal conductivity
 of spin-1/2 chains \cite{kane}, and it could be very interesting to compare this to numerical results.
Here, results  for the frequency 
 dependence of $\kappa$ are presented for the case of {\em off-diagonal} disorder at finite temperatures. 
 
\begin{figure}[!ht]
\begin{center}
\includegraphics[width=0.45\textwidth]{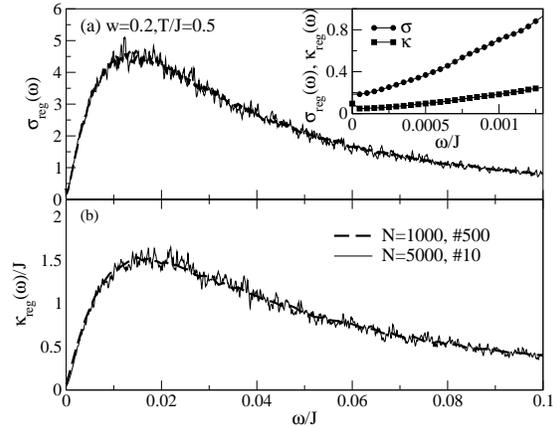}
\end{center}
\caption{Spin [panel (a)] and thermal conductivity [panel (b)] of an $XX$ chain with a Gaussian distribution of
random couplings $J_l$, centered around $J=1$ and with a width of $w=0.2$ (see text).
Curves for $N=1000$(5000) sites and 500(10) random realizations for the same parameters are displayed. An imaginary broadening of
$10^{-4}$ is used. The inset shows a blow-up of the low frequency region for the larger system size. }
\label{fig1}
\end{figure}

The Hamiltonian  in terms of spin-$1/2$ operators $S_l^{\pm,z}$ acting on site
$l$ and
with periodic boundary conditions (PBC) reads:
\begin{equation}
H= \frac{1}{2}\sum_{l=1}^N \, J_l \, (S_l^+ S_{l+1}^- + h.c.)  \,,
\label{eq1}
\end{equation}
where $N$ is the number of sites. It  can equivalently be written in terms of fermionic operators $c_l^{\dagger}$:%\cite{lieb61}:
\begin{equation}
H= \frac{1}{2}\sum_{l=1}^{N} \, J_l \, (c_l^{\dagger} c_{l+1} + h.c.) \,.
%+ (e^{i\pi\mathcal{N}}/2)(c_{1}^{\dagger}c_{N} +h.c.)\,.
\label{eq2}
\end{equation}
In principle,  a subtlety arises since the PBCs for the spin operators 
give rise to a nontrivial boundary term
for the fermions, which depends on the number $\mathcal{N}$ of fermions \cite{rieger}. This dependence,
however, is neglected here, and PBC are imposed for all $\mathcal{N}$, since
boundary effects can be expected to vanish for large system sizes.\\\indent
Introducing a spinor $\psi^{\dagger}=(c_{1}^{\dagger},\dots,c_{N}^{\dagger})$, the Hamiltonian
can be written as $H= \psi^{\dagger} \mathcal{A} \psi$, where $\mathcal{A}$ is a symmetric
$N\times N$  band matrix, with nonzero elements $A_{l,l+1}=J_l/2$ and $A_{1,N}=J_1/2$. 
Randomness in the $J_l$ is therefore called off-diagonal disorder, while a spatially varying
magnetic field realizes {\it diagonal} disorder.\\\indent
While a transformation to momentum space diagonalizes $H$
for the translationally invariant case ($J_l = J$),
the computation of single-particle eigenenergies is still straightforward for random couplings
(see, e.g., \cite{rieger}). By means of a unitary transformation $\mathcal{U}$ with
 $c_l=\sum_{\mu}\mathcal{U}_{l\mu}\eta_{\mu}$  that diagonalizes the
 matrix $\mathcal{A}$, the Hamiltonian can be written as
 %\begin{equation}
$H = \sum_{\mu} \epsilon_{\mu} \eta_{\mu}^{\dagger}\eta_{\mu}^{ } \label{eq:hmu}\,,$
%\label{eq3}
%\end{equation}
where $\epsilon_{\mu}$ are the single-particle eigenenergies.\\\indent
The energy current operator corresponding to Eq.~(\ref{eq2}) is
$\mathcal{J}=(i/4)\sum_l(c_l^{\dagger}c_{l+2}-h.c.)=i\sum_{\mu\nu}\eta_{\nu}^{\dagger} \mathcal{J}_{\nu\mu}
\eta_{\mu}$.
Within linear response theory,
$\kappa_{\mathrm{reg}}(\omega)$ is given by \cite{kramer93}:
  \begin{equation}
\kappa_{\mathrm{reg}}(\omega)= \frac{\beta}{\omega}
\sum_{\epsilon_{\mu}\not= \epsilon_{\nu}} \-\-|\mathcal{J}_{\nu\mu}|^2 \, 
\left[ f(\epsilon_{\mu})-f(\epsilon_{\nu})\right] \,\delta(\omega-\Delta \epsilon)\,,
\label{eq4}
\end{equation}
where $f(\epsilon)=1/(\mathrm{exp(\beta\epsilon)}+1)$ denotes the Fermi-function,
$\beta=1/T$, and $\Delta \epsilon=\epsilon_{\nu}-\epsilon_{\mu}$.
\\\indent
As an example, a Gaussian distribution of random couplings is considered:
$P(J_l) \propto e^{-(J_l-J)^2/w^2}$, with $J=1$ and $w=0.2$. The  choice of
the distribution does not affect the results, quantitatively consistent results
for $\kappa(\omega)$ are obtained with other $P_l$ (e.g. box or binary
distribution) by fixing the first moments
of the distribution function. 
Figure \ref{fig1} shows results for $\kappa_{\mathrm{reg}}(\omega)$ [panel(b)] and, for comparison,
$\sigma_{\mathrm{reg}}(\omega)$ [panel(a)], which is the  spin conductivity corresponding to 
the current operator $\mathcal{J}_{\mathrm{s}}=(i/2)\sum_l(c_l^{\dagger}c_{l+1}-h.c.)$.
The parameters are $T/J=0.5$ and $N=1000,5000$. An imaginary broadening of $10^{-4}$ was used.
The overall form of both quantities is quite similar and resembles that known
for $\sigma_{\mathrm{reg}}(\omega)$ \cite{motrunich}. Moreover, both curves 
exhibit a maximum at roughly the same frequency ($\omega/J\approx 0.015$ for the
parameters of Fig.~\ref{fig1}). Both finite-size effects and statistical fluctuations are
small in the low-frequency limit.\\\indent
The behavior of $\kappa_{\mathrm{reg}}(\omega)$ at low frequencies 
is of particular interest as it determines the dc-conductivity (see Refs.~\cite{zotos04} for 
the case of clean spin systems). While the Drude weights vanish in the presence of disorder,
the results of this work indicate finite dc-conductivities for both spin and thermal 
transport at finite temperatures. This is illustrated in the inset of Fig.~\ref{fig1}.
A more detailed analysis of the finite-size scaling as well as the temperature dependence
will be presented elsewhere.
 %The transport coefficient is evaluated by means of linear response 
 %theory\cite{mahan}:
 %\begin{equation}
 %\kappa(\omega)  
%= \frac{\beta}{N}
% \int_0^{\infty} dt\, e^{i(-\omega+i0^+ )t}
 %        \int_0^{\beta} d\tau \langle j_{\mathrm{th}}\,j_{\mathrm{th}}(t+i\tau)\rangle \, ,
%\end{equation}
% where $\beta=1/T$ is the inverse temperature, $j_{\mathrm{th}}$ denotes the thermal current operator

%\section*{Acknowledgment}
The author acknowledges fruitful discussions with W. Brenig, B. B\"uchner, C. Hess, and A. Honecker. This work was 
supported by the Deutsche Forschungsgemeinschaft through SPP 1073 and grant
HE-5242/1-1.


\begin{thebibliography}{99}

\bibitem{exp} See, e.g.: A.~V. Sologubenko et al.,
Phys.\ Rev.\ Lett. {\bf 84} (2000) 2714; C.~Hess et al., Phys.\ Rev.\ B {\bf 64} (2001) 184305; K.~Kudo et al., J. Phys. Soc. Jpn. {\bf 70} (2001) 437.
\bibitem{zotos_rev} See: X. Zotos and P. Prelov{\v{s}}ek: in {\it  Strong Interactions in Low Dimensions}, Kluwer
Academic Publishers, 2004; and further references therein.
\bibitem{zotos97} X.~Zotos, F.~Naef, and  P.~Prelov{\v{s}}ek, Phys.\ Rev.\ B {\bf 55} (1997) 11029.
\bibitem{xxz} A.~Kl\"umper and  K.~Sakai, J. Phys. A {\bf 35} (2002) 2173;  K.~Sakai and A.~Kl\"umper 
J. Phys. A {\bf 36} (2003) 11617; F.~Heidrich-Meisner et al., Phys.\ Rev.\ B {\bf 66} (2002) 140406; 
{\it ibid.} {\bf 68} (2003) 134436; {\it ibid.} {\bf 71} (2005) 184415.
\bibitem{mahan} G. D. Mahan, {Many-{P}article {P}hysics}, Plenum Press, New York London, 1990.
\bibitem{kramer93} B. Kramer and A. MacKinnon, Rep. Prog. Phys. {\bf 56} (1993) 1469.
\bibitem{kane}
C.~I. Kane and M.~P.~A. Fisher,
\newblock Phys.\ Rev.\ Lett. {\bf 76} (1996) 3192; A.~V. Rozhkov and A.~L. Chernyshev,
\newblock Phys.\ Rev.\ Lett. {\bf 94} (2005) 087201.

%\bibitem{lieb61} E. Lieb, T. Schulz, and D. Mattis, Ann. Phys. (N.Y.) {\bf 16} (1961) 407.
\bibitem{rieger} N. Laflorencie and H. Rieger, Eur. Phys. J. B {\bf 40} (2004) 201.
%\bibitem{kluemper02} A.~Kl\"umper and  K.~Sakai, J. Phys. A {\bf 35} (2002) 2173. 
%\bibitem{alvarez02} J.~V. Alvarez and  C.~Gros, Phys.\ Rev.\ Lett. {\bf 88} (2002) 077203.
%\bibitem{shimshoni03} E.~Shimshoni, N.~Andrei und  A.~Rosch, Phys. Rev. B {\bf 68} (2003) 104401.
%\bibitem{hm02} F.~Heidrich-Meisner et al., Phys.\ Rev.\ B {\bf 66} (2002) 140406; {\it ibid.} {\bf 68} (2003) 134436; Phys. Rev. Lett. {\bf 92} (2004) 069703.
%\bibitem{orignac03} E.~Orignac, R.~Chitra, and  R.~Citro, Phys.\ Rev.\ B {\bf 67} (2003) 134426.
%\bibitem{saito03} K.~Saito, Phys. Rev. B {\bf 67} (2003) 064410.
\bibitem{motrunich} O.~Motrunich, K.~Damle, and D.~A. Huse, Phys. Rev. B {\bf 63} (2001) 134424.
\bibitem{zotos04} X.~Zotos, Phys. Rev. Lett. {\bf 92} (2004) 067202; 
F.~Heidrich-Meisner et al., Physica B {\bf 359-361} (2005) 1394.
%\bibitem{hm05} F.~Heidrich-Meisner et al., Physica B {\bf 359-361} (2005) 1394.




\end{thebibliography}
\end{document}